.

# Spin-Chain Incipient Magnetocaloric Effect and Rare-Earth(R) Controlled Switching in the Haldane-Chain System, $R_2BaNiO_5$

**Mohit Kumar,[1] Gourab Roy,[1] Sayan Ghosh,[1] Ekta Kushwaha,[1] Kiran Singh,[2] and Tathamay Basu*[1]**

*[1] Rajiv Gandhi Institute of Petroleum Technology, Jais, Amethi, 229304, Uttar Pradesh, India*

*[2] Department of Physics, Dr. B. R. Ambedkar National Institute of Technology, Jalandhar 144008, India*

**** Corresponding author: tathamay.basu@rgipt.ac.in***

## Abstract

We have experimentally investigated the magnetocaloric effect (MCE) of a prototype spin-frustrated one-dimensional spin-chain system, the famous Haldane-chain system, $R_2BaNiO_5$ (R = Dy, Nd, Gd, Er). The significant MCE is observed far above long-range ordering, which is attributed to the change in magnetic entropy due to short-range spin correlation arising from (low-dimensional) magnetic frustration. Such a spin-chain incipient MCE above long-range ordering is rarely reported. Interestingly, multiple magnetocaloric switching from conventional to inverse MCE (and vice versa) are observed below long-range magnetic ordering, as a function of temperature and magnetic field, for the R = Nd, Dy, and Er members. However, such MCE switching is absent in the Gd member, which is an S-state atom (orbital moment L = 0). Our systematic investigation of this series suggests that the interplay between crystal-electric field (CEF), strong spin-orbit coupling (SOC) and rare earth anisotropy of R-ions play an important role in spin reorientation, which leads to multiple MCE switching due to intriguing changes in magnetic and lattice entropy. The maximum change of entropy for Er, Gd, Dy and Nd is 7.8, 6.8, 4.0 and 1.0 J $Kg^{-1}$ $K^{-1}$ respectively. Our study presents a pathway for tuning MCE switching and the MCE effect over large temperature regions in d-f coupled spin-frustrated and spin-chain oxide systems.

## 1. Introduction

The magnetocaloric (MC) material, used in adiabatic demagnetization refrigeration (ADR) devices, is in high demand because of its environment-friendly application for energy-efficient refrigeration and cryogenic cooling (down to sub-Kelvin temperature) without employing pricey Helium, which is useful in space-science and fuel industries [1-5]. Fundamentally, the change in an external magnetic field (*H*) causes an alteration in the temperatures (cooling) in some magnetic materials which is called magnetocaloric effect (MCE) [6-8]. The schematic diagram of MCE is shown in Fig.1. When an external magnetic field is applied to an MC material in isothermal condition, its magnetic entropy decreases as the spins get ordered, if, the applied field is removed adiabatically, the systems become disordered due to this adiabatic demagnetization, and the material cools down adjusting the increase of magnetic entropy through lattice [9,10]. These thermodynamic processes and responses are usually reversible, which leads to energy-efficient real-world applications. The MCE-based magnetic refrigerators have the potential to save up to 30% of energy consumption compared to other conventional refrigeration procedures and do not require the use of any gases which is highly important for a clean environment [11]. The detection of a substantial MCE at room temperature in an intermetallic compound, $Gd_5(Si_2Ge_2)$, [12] near its first-order phase transition has sparked considerable interest within the research community to obtain efficient MC materials for potential applications [13]. The compound containing heavy rare earth elements is excellent for producing substantial MCE due to its large magnetic moment [14,15,16]. The frustrated magnets, having more than one ground state and thereby creating disorder in the spin arrangement, are considered good candidates to achieve large MCE and cool down the system from a few Kelvin to sub-Kelvin. Ever since the remarkable discovery of a massive MCE in $Gd_5Si_2Ge_2$, there has been a growing interest in studying materials that undergo magnetic phase transitions and/or structural transitions under the influence of an external field [12, 17-19]. However, one of the main drawbacks of the reported MCE materials is that they show large MCE around the magnetic phase transition, which sharply decays below magnetic ordering [15,16,20]. It is challenging to obtain an MCE with a large temperature window. In this manuscript, we have addressed this issue documenting the MCE over a large temperature window.

In recent years, the exploration of magnetocaloric materials has expanded to include frustrated oxide systems. Notable examples include multiferroic compounds such as $RMnO_3$[21] (R = rare earth), $RMn_2O_5$,[22] and Haldane-type systems like $R_2BaZnO_5$ [23]and $R_2BaCuO_5$,[24] along with other related compounds, for example $NaYbGeO_4$ [25] $RE_3Co_2Ge_4$ [26]and $Ba_3RRu_2O_9$ (R=Ho, Gd, Tb and Nd) [27] etc [15,28,29]. The strong magnetic frustration (arising from geometry or/and competing R(4f) and Mn(3d) spins), and spin-lattice coupling in this multiferroic system favor large change in magnetic entropy. The multiferroic compound $TbMn_2O_5$ exhibits a significant change in magnetic entropy and shows a large MCE around magnetic ordering [30]. Another multiferroic compound $GdCrTiO_5$ also exhibits large conventional MCE at very low temperatures, probably due to exchange frustration from Gd(4f)-Cr(3d) spins [31-32]. The Haldane spin-chain system $R_2BaNiO_5$ has been attracted for over 30 years, due to their unique properties associated with fascinating phenomena, such as the unusual spin-gap (Haldane gap for S=1 spin-system) arising from Ni-spin in magnetically ordered state, multiferroicity and strong magnetoelectric coupling [33]. The large magnetic moment of heavy rare-earth and strong spin frustration arising from various competing exchange interactions makes this system suitable for exploring the magnetocaloric effect. However, no detailed study of MCE on the $R_2BaNiO_5$ series exists. Here we have explored the MCE of the $R_2BaNiO_5$ series and demonstrate the role of different rare-earth on MCE. We observe MCE switching for R=Dy, Er, and Nd, which is absent for Gd. We have documented significant MCE even above long-range magnetic ordering, which is attributed to short-range magnetic correlation from spin-chain in this Haldane-chain system.

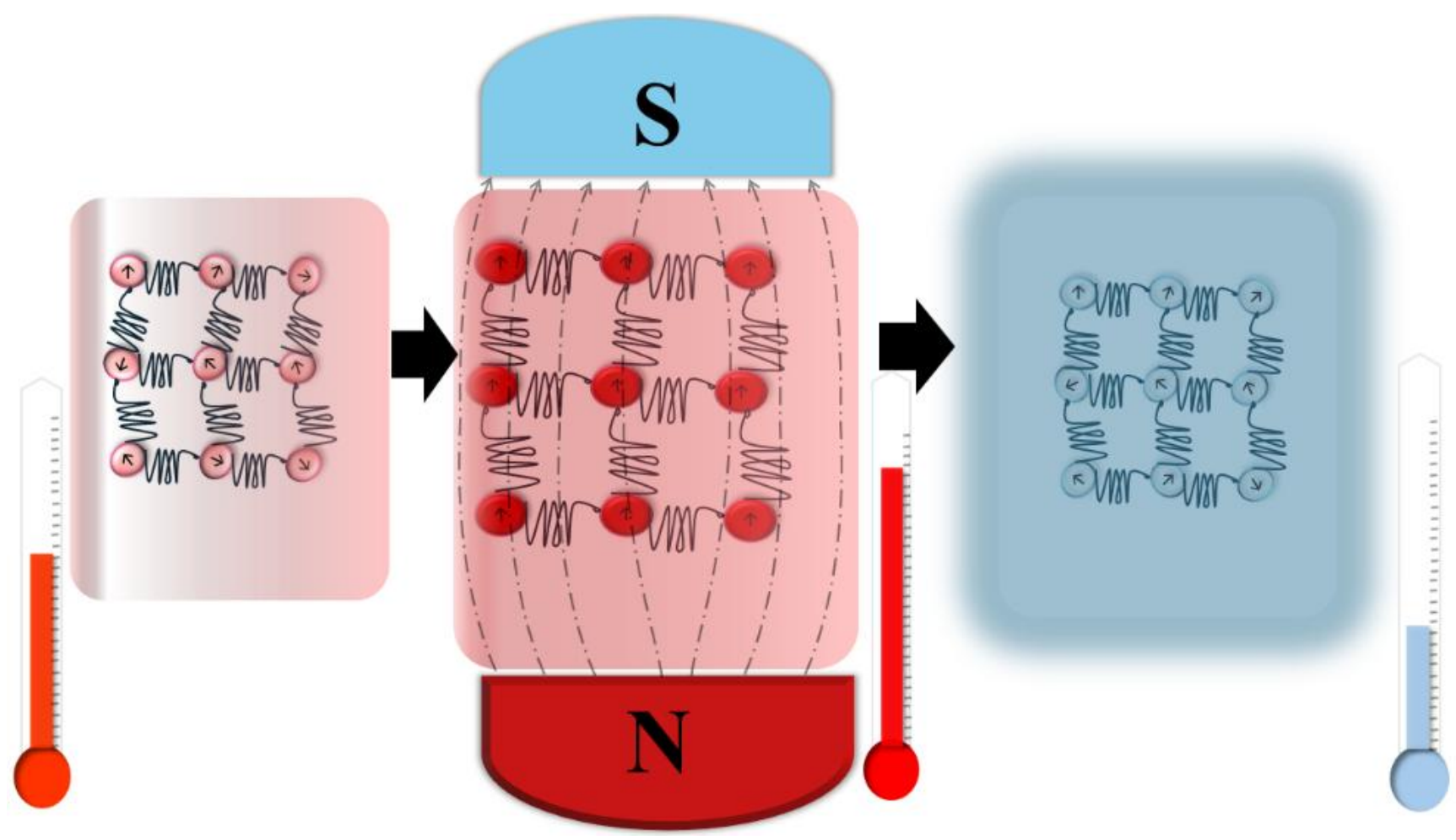

**Figure 1 Schematic diagram of Magnetocaloric effect (MCE).**

## 2. Experimental Details

The polycrystalline sample of $R_2BaNiO_5$ was prepared by a standard solid-state reaction using mixtures of high purity precursors: $BaCO_3$(Sigma, >99.8%), NiO(Sigma, 99.9%), and $R_2O_3$ (Sigma, 99.99%) (R=Dy, Er, Nd, and Gd) which were mixed in an agate mortar and pestle and pressed into pellets, and finally sintered at 1200 °C with several heating as described in Ref. [34]. Further, the phase purity is confirmed by the powder X-ray diffraction (see Fig.S1 in Supplementary Information (SI)). The heat capacity (C) was measured by a physical property measurement system (PPMS, Quantum Design) using relaxation method at various field with a temperature range of 1.8-80 K. The temperature DC magnetization was carried out using superconducting quantum interference device (SQUID) and isothermal magnetization as a function of magnetic field was performed employing vibrating sample magnetometer (VSM), procured from Quantum Design. The preliminary magnetic results are depicted in S.I.

## 3. Results:

The compound $R_2BaNiO_5$ crystallizes in an orthorhombic structure (space group *Immm*), consisting of distorted $NiO_6$-octahedra along the chain (crystallographic a-axis). The chains are connected by $RO_6$-octahedra, which develops long-range antiferromagnetic (AFM) ordering through Ni-O-R-O-Ni superexchange interactions. The magnetic structure is intriguing for different R-members of this series. The size of the rare-earth ions significantly affects the structural parameters, which leads to geometric distortions in the structure ($NiO_6$-octahedral distortion) [35]. The crystal-electric field (CEF) and spin-orbit coupling (SOC) of rare-earth ions cause a significant effect on the magnetic ground state, which yields intriguing magnetic behavior for various R-members in this series [36-40].

### *3.1. MCE in $Dy_2BaNiO_5$:*

The compound $Dy_2BaNiO_5$ orders at 58 K ($T_N$) followed by another magnetic feature around 35 K ($T_{max}$) associated with spin reorientation and a distinct feature below 12K [36]. The heat capacity (*C*) as a function of temperature (*T)* for various magnetic fields (*H*=0-14T) for $Dy_2BaNiO_5$ is shown in Fig.2a. The λ-like peak (though weak) at 58 K in the absence of magnetic field confirms the magnetic phase transition. The magnetic peak temperature ($T_N$) decreases and broadens with increasing magnetic field, indicating the antiferromagnetic ordering of this compound. Interestingly, with decreasing temperature, a substantial change in the absolute value in *C(T)* is observed below 40 K down to 15 K, suggesting a clear continuous change in the magnetic entropy. This could result from spin reorientation and increment of moment value by lowering the temperature in this system. No additional feature is observed down to 2 K for the absence of a magnetic field. The *C(T)* curve for H=0 nearly superimposed with the feature in the presence of magnetic field up to 3 T in low-*T* region (see inset of Fig. 2a). However, a clear peak ~ 8.5 K appears under application of a very high magnetic field of 5 T (inset of Fig. 2a). This peak shifts to higher temperature with increasing magnetic field (for instance, 11.5 K for 7 T) which is suggestive to ferromagnetic nature of this system. This peak position in *C(T)* remains nearly the same, however, the value gets suppressed under the application of a further higher magnetic field (say, 9T and 14 T), indicating a near saturation of the spin moments. We have calculated change in entropy of the system with the different magnetic fields using the formula,

$$\Delta C_p(T)=C_p\,(T, H) - C_p(T,0)$$

$$\Delta S_m = \int_{T_o}^{T} \frac{\Delta C_p(T')}{T'}\, dT'$$

The change in magnetic entropy ($S_M$) with respect to $S$ ($H$=0) is plotted in Fig. 2b. This system exhibits a coexistence of conventional and inverse MCE, with entropy changes of 4 J $Kg^{-1}$ $K^{-1}$ and -3.1 J $Kg^{-1}$ $K^{-1}$, respectively. Notably, it demonstrates remarkable switching between these effects, which can be tuned by varying temperature and magnetic field. The magnetic entropy for $H$=3 T shows negligible changes below $T_N$ down to 30 K. However, a significant gradual enhancement of magnetic entropy is observed under application of a high magnetic field in this temperature region (say 2 J/Kg-K for 14 T). An earlier neutron diffraction study reported [35] that $Ni^{+2}$-spin moments reorient from the +*a*-axis to the +*c*-axis towards $Dy^{+3}$-moments below $T_N$ with decreasing temperature and the magnetic moments of Ni and Dy slowly go towards saturation with decreasing temperature. The application of the magnetic field will have a pronounced effect on the spin reorientations, on which the spins will try to orient towards the application of the magnetic field, manifested as a large change of magnetic entropy in this region. With the further lowering of the temperature below 30 K, the $S_M$ value switches from +Ve to -Ve, which indicates an inverse MCE in the presence of a magnetic field [27, 38]. The *T*-dependent magnetic susceptibility shows a clear broad peak around 35 K ($T_{max}$) (see SI Fig.S2), assigned to the alignment of Ni-spins towards Dy-moments and saturation of Ni-moments. We observe the switching in MCE at similar temperatures. In this compound, various magnetic exchange interactions are competing, e.g., Dy-O-Ni super-exchange, Ni-O-Ni super-exchange, and Dy-Dy direct exchange, which makes the system magnetically frustrated (exchange-frustration). Two *H*-induced meta-magnetic transitions, one at ~ 4.5 T ($H_{C1}$) and another at 6.5 T($H_{C2}$), were reported in isothermal magnetization (see SI Fig.S6), which was assigned to a magnetic phase transition [36, 39]. The appearance of a peak ~ 8.5 K in *C(T)* under *H*=5 T (H>$H_{C1}$) is consistent with this phase transition associated with $H_{C1}$. The large negative change in $S_M$ for 5T (Fig.2b) is attributed to the change in spin pattern around $H_{C1}$, which causes a large change in magnetic entropy. Most likely the spin-pattern turns into canted AFM from AFM-structure. The application of higher magnetic field of 7T (H>$H_{C2}$) yields a more ferromagnetic like spin-pattern as indicated by heat-capacity [Fig.2a] and isothermal magnetization [36]. Most probably, an application of a high magnetic field flops the spins towards the magnetic field and enhances the ferromagnetic contribution. This $2^{nd}$ magnetic phase transition produces the large entropy in the same direction as depicted in Fig.2b. Therefore, the large change in $S_M$ for 5 T and 7 T is related to order-order phase transitions from one magnetic ground state to another magnetic ground state due to spin-flop. The isothermal M(H) results reveal that the application of further magnetic field ($H_{C2}$< *H* <14 T) does not show any signature of magnetic phase transition; however, the moment gets enhanced towards saturation with increasing magnetic field and nearly gets saturated at 14 T. The magnetic entropy above $H_{C2}$ (e.g. 9 T and 14 T) starts to decrease compared to its earlier value at 7 T, the positive change in entropy from its earlier value is attributed to the saturation of the spin-moments in this ordered state.

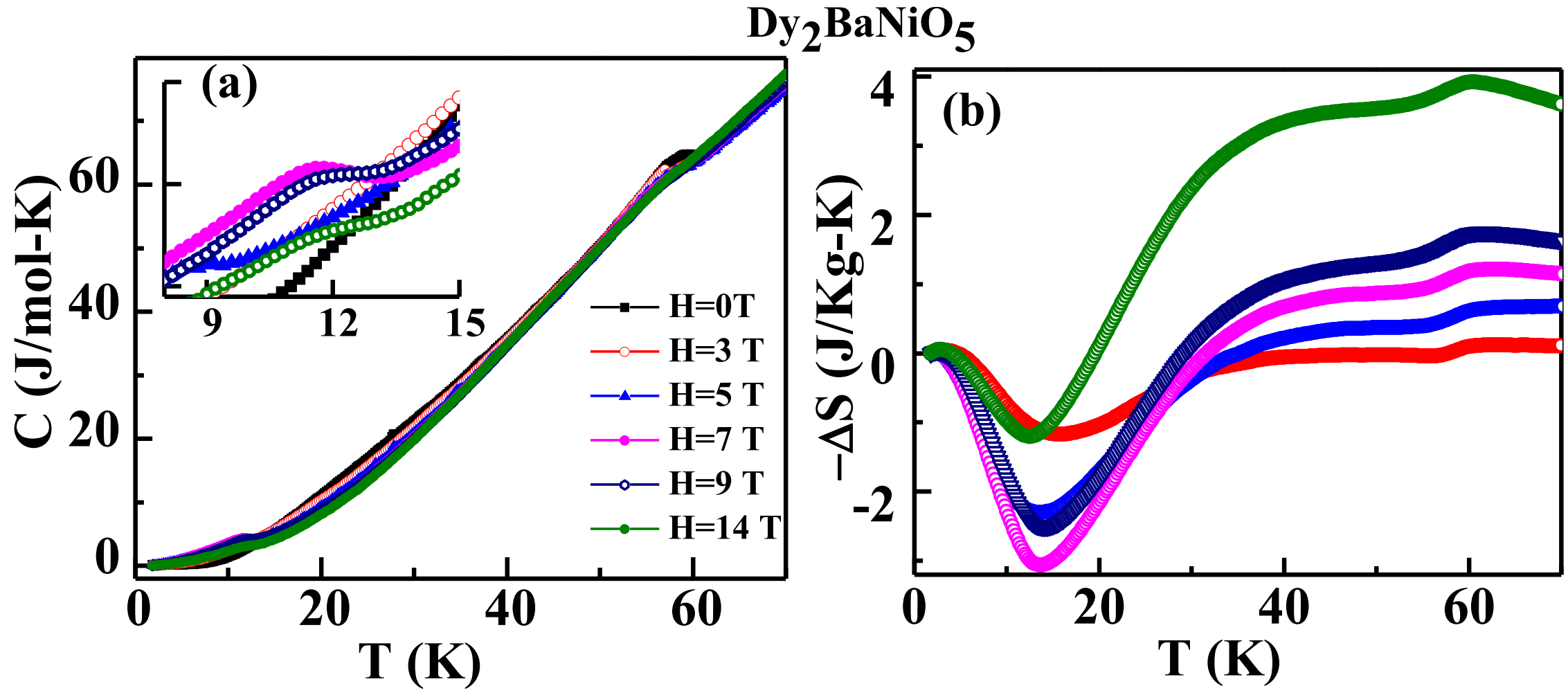

**Figure 2: (a) Temperature dependence of specific heat; the inset shows an enlarged view around the transition temperature. (b) Magnetic entropy change (ΔS) as a function of temperature under various applied magnetic fields for $Dy_2BaNiO_5$.**

Interestingly, the MCE persists even above $T_N$. The value of MCE for 14 T at T=70 K is ~ 4 J/Kg-K, which is nearly the same value below 58 K ($T_N$), in contrast to most of the MC materials, which show a sharp drop of MCE above $T_N$ [12,15,16,29].

### *3.2. MCE in $Nd_2BaNiO_5$*

Now, we will discuss the MCE of a light-rare-earth member in this family, $Nd_2BaNiO_5$. The temperature-dependent heat-capacity and the calculated change in entropy under different magnetic fields is shown in Fig.3a and 3b, respectively, for the Haldane chain system $Nd_2BaNiO_5$. The neodymium members show antiferromagnetic ordering at $T_N$ = 48 K [40]. Two additional magnetic anomalies are reported as also evident from magnetic susceptibility data, one broad magnetic feature ~26-30 K and another below 10 K [39, 40]. The $Nd_2BaNiO_5$ Haldane chain system exhibits a significant inverse MCE below 26 K, which subsequently switches to a conventional MCE above 26 K. The magnitude of the inverse MCE, indicated by the change in magnetic entropy, increases with increasing magnetic field. The maximum change in magnetic entropy for the inverse MCE is -2.4 J $Kg^{-1}$ $K^{-1}$ around 12 K, while for the conventional magnetocaloric effect, it is 1 J $Kg^{-1}$ $K^{-1}$ around 47 K under an applied magnetic field of 14 T. The Ni moments start to order antiferromagnetically below 48 K and gradually achieve saturation below 35 K [40]. The Nd moments simultaneously order at 48 K; however, the magnitude of its moment increases relatively slowly with decreasing T, achieving a saturation down to 5 K [40]. The Unlike Dy-member, no spin-reorientation is observed.

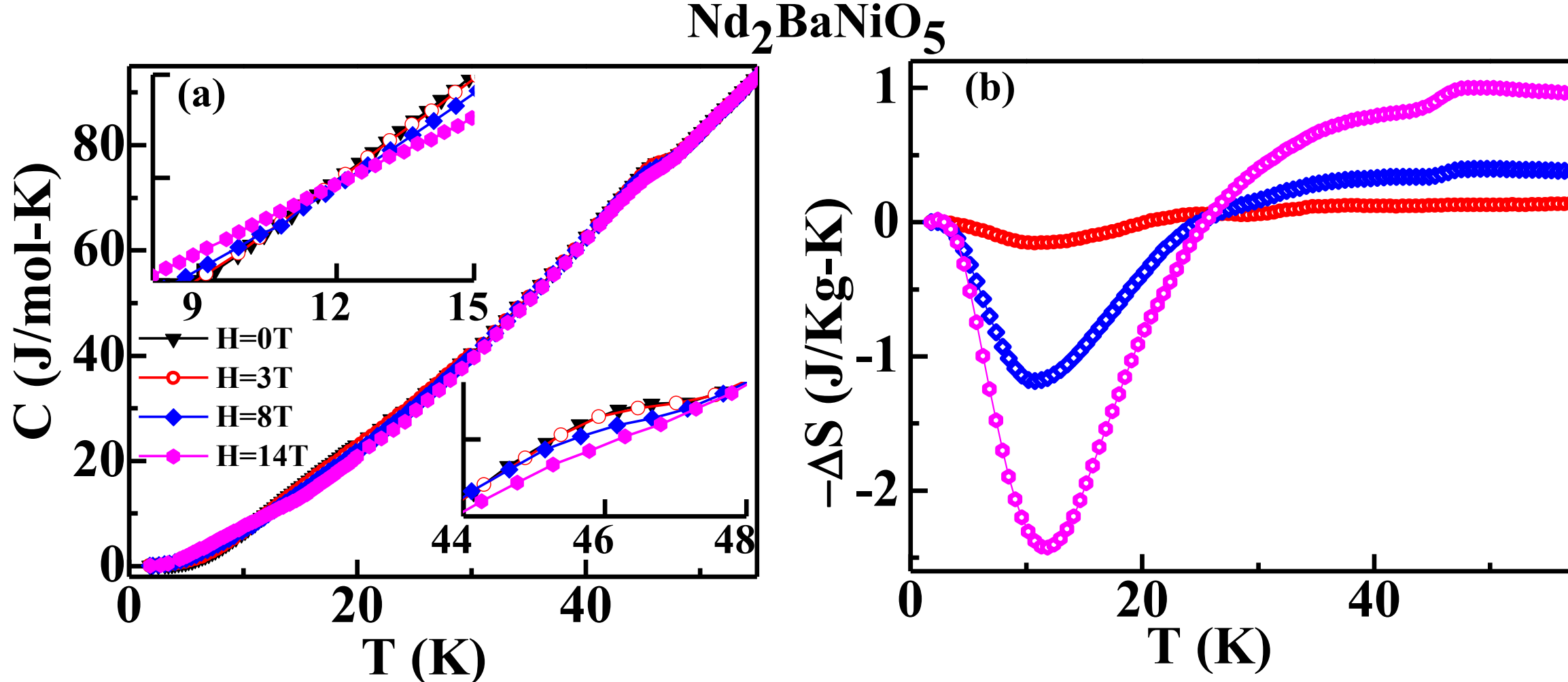

**Figure 3(a) Temperature dependence of specific heat; the inset shows an enlarged view around the transition temperature. (b) Magnetic entropy change (ΔS) as a function of temperature under various applied magnetic fields for $Nd_2BaNiO_5$.**

The broad magnetic anomaly around 26-30 K in magnetic susceptibility in SI Fig.S3 is assigned to a change in the magnetic ground state from its low-temperature state, due to the population of the higher-level Nd-Kramer's doublet at 25 K, which is split due to the Nd-Ni exchange interaction [40, 41]. The cross-over of MCE at 26 K is assigned to change the magnetic state, from one ordered state to another ordered state. At further lower temperatures below 10 K, the $S_M$ exhibits a dip, corresponding to the third anomaly, which was ascribed to a spin-glass-type feature in an earlier report [39]. The value of $S_M$ at this anomaly, that is, the height of dip, gradually increases with increasing magnetic field. This indicates an enhancement in entropy with the applied field, arising from a transition to a more disordered magnetic structure (glassy type) as the temperature decreases. Similar to all other heavy-rare-earth, the light-rare-earth member, $Nd_2BaNiO_5$, also exhibits MCE far above long-range ordering. This indicates that the MCE above $T_N$ is a characteristic of this family.

### *3.3. MCE in $Gd_2BaNiO_5$*

Now we will discuss MCE of another heavy rare-earth member, $Gd_2BaNiO_5$. The gadolinium is a S-state atom containing zero orbital moment, which makes $Gd^{+3}$ ($f^{\,7}$) unique in the

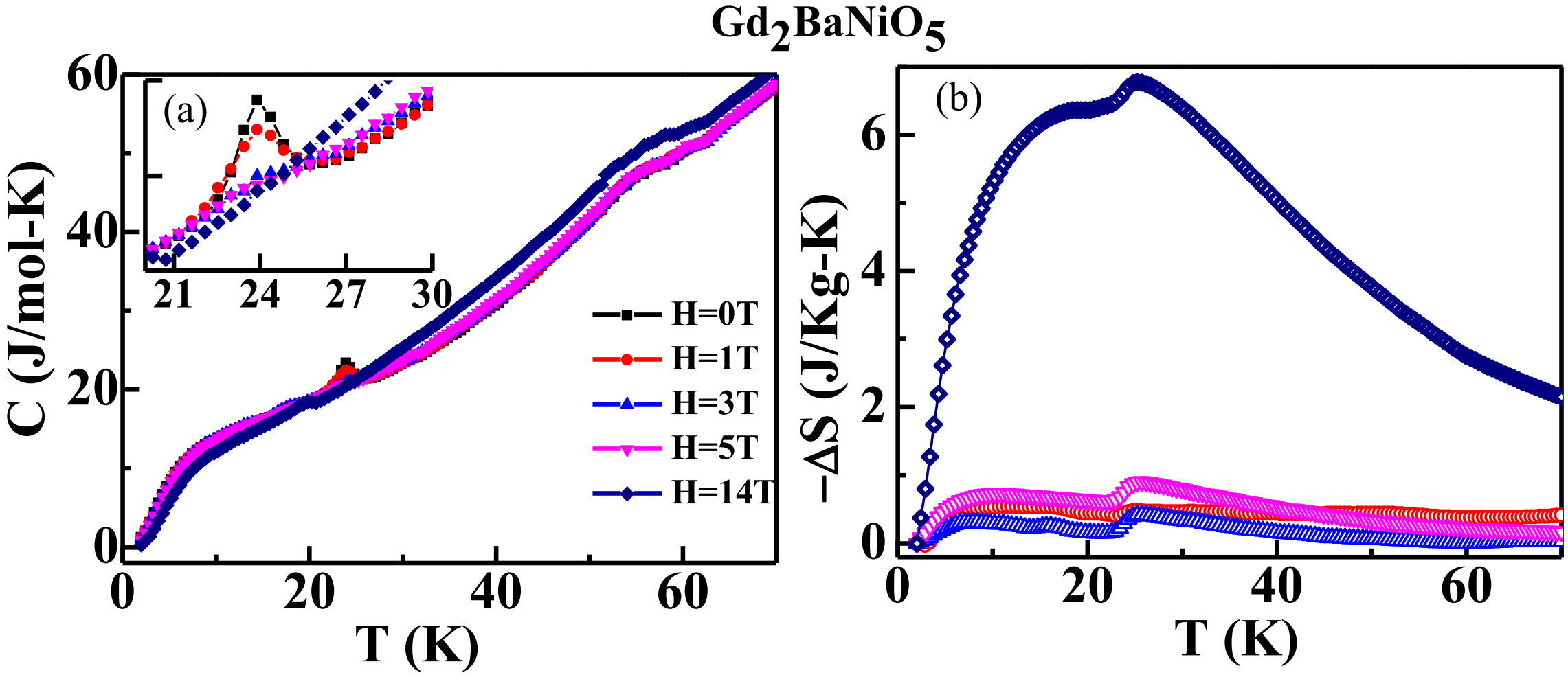

**Figure 3: (a) Temperature dependence of specific heat; the inset shows an enlarged view around the transition temperature. (b) Magnetic entropy change (ΔS) as a function of temperature under various applied magnetic fields for $Gd_2BaNiO_5$.**

lanthanide/ rare-earth family.

Fig.4a presents the heat capacity as a function of temperature for $Gd_2BaNiO_5$, both with and without an applied magnetic field. The interactions mediated by Gd between the nickel S=1 chains result in antiferromagnetic ordering, which emerges at $T_N$= 55 K. Additionally, another magnetic anomaly is observed at temperatures above 25 K which is also observe in magnetic susceptibility as show in Fig.S4. Fig.4b illustrates the calculated change in entropy of the system under different magnetic fields. This gadolinium-based Haldane chain system exhibits the conventional magnetocaloric effect, in contrast to other heavy rare-earth members, Dy and Er. The largest entropy change is 6.8 J $Kg^{-1}$ $K^{-1}$ for an applied magnetic field of 14 T around 25 K. The Ni-Gd super-exchange interaction is dominating in this T-region. The application of a magnetic field tries to destroy the spin arrangement and Ni and Gd spins tend to orient along the H, which might cause a large enhancement of positive $S_M$ (conventional MCE) due to the disorder arrangement of the spin-ordered structure.

### *3.4. MCE in $Er_2BaNiO_5$*

The compound $Er_2BaNiO_5$ magnetically orders at 32 K ($T_N$) where Er and Ni start to order, followed by another broad magnetic feature in magnetic susceptibility around 16-20 K (see SI Fig.S5) which is attributed to the splitting of the Kramer's doublet of trivalent Er by the exchange interaction [43] and related to spin-orientation [39]. Fig.5a shows the heat capacity as a function of temperature under various magnetic fields (H=0-14 T]. It is already documented in past literature [39,42] that the magnetic long-range ordering at 32K is weakly traced in magnetic susceptibility and heat capacity as the moment of Er and Ni is very small below $T_N$ and a gradual slow change of moment from high to low temperature does not yield any sharp feature around ordering. A cross-over of C(T) for various magnetic fields is observed ~ 32 K [Fig.5a]. With further decreasing temperature, the C(T) starts to increase below 20 K developing a broad peak ~ 10 K for H=0 and 1 T. The application of a higher magnetic field shifts this broad feature to lower temperatures, implying the dominance of AFM interactions in this system. The change in the position of the peak temperature with increasing magnetic field in $Er_2BaNiO_5$ is in sharp contrast with $Dy_2BaNiO_5$. The calculated magnetic entropy is plotted in Fig.5b as a function of temperature from 2-70 K for various magnetic fields. Starting from 2 K, the $S_M$ initially decreases with increasing temperature, showing a dip around 4-6 K, and further enhances with warming the temperature. A weak magnetic phase transition is predicted around 4 K in our earlier report [ 35]. The dip in $S_M$ occurs at this anomaly. This change in entropy further supports the magnetic phase transition. The Er-Er interaction dominates at very low T ($<5$ K), whereas Ni-Er exchange interaction dominates at higher temperatures above 10 K.

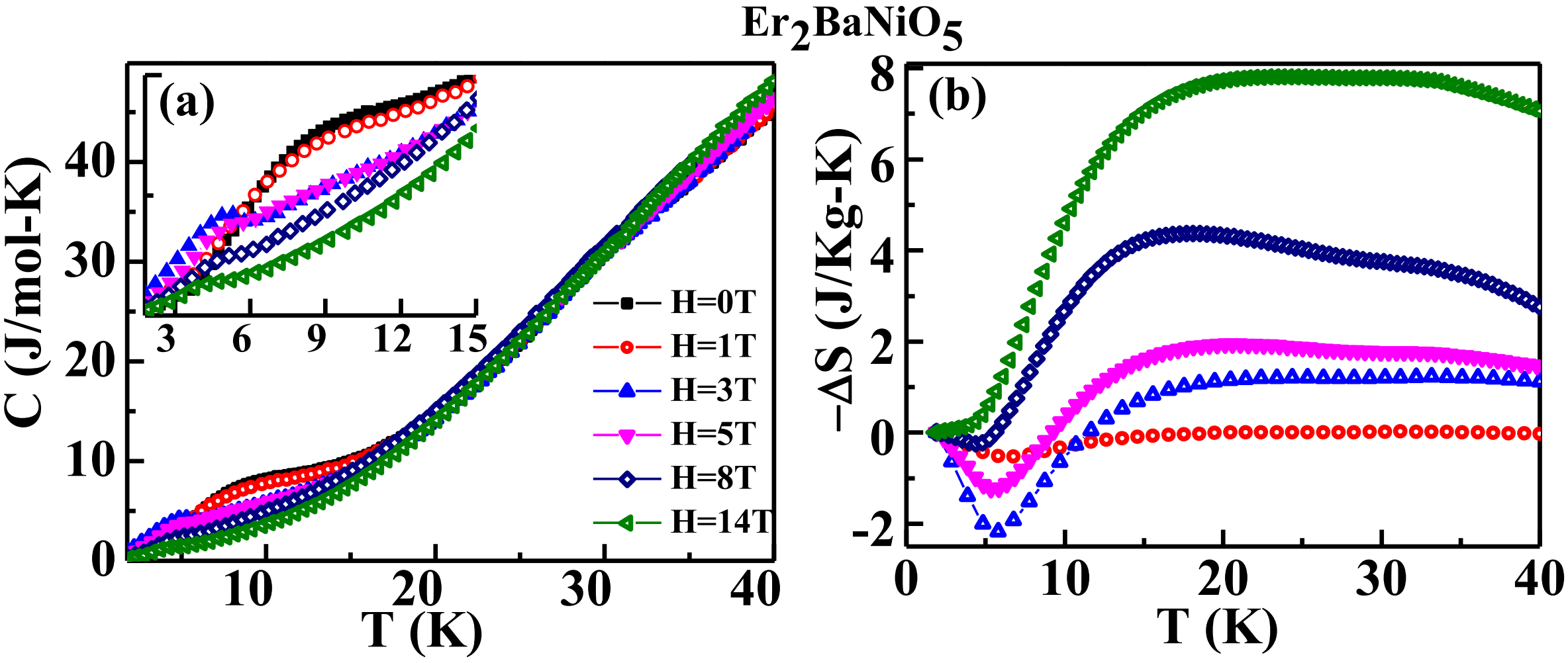

**Figure 5: (a) Temperature dependence of specific heat; the inset shows an enlarged view around the transition temperature. (b) Magnetic entropy change (ΔS) as a function of temperature under various applied magnetic fields for $Er_2BaNiO_5$.**

Therefore, the dominance of Er-Er exchange interaction over Ni-Er exchange interaction could be one possible reason for a change in magnetic phase (spin-arrangement in this ordered state), and thus, a significant change in $S_M$ around 5 K. The nature of $S_M$ remains the same under the application of a large magnetic field of 14 T. However, the entropy changes of 7.8 J $Kg^{-1}$ $K^{-1}$ under the application of a 14 T magnetic field and -2.2 J $Kg^{-1}$ $K^{-1}$ for an application of a 3 T magnetic field. The enhancement of the magnetic field favors the saturation of magnetic moment which helps in the enhancement of $S_M$. Interestingly, we observe magnetic entropy even above magnetic ordering. Eventually, significant $S_M$ exists at 90 K, far above $T_N$.

## 4. Crystal Electric Field (CEF) Excitation: Theoretical Calculation

To interpret the crystal electric field (CEF)excitations theoretically, we employ point charge calculations. In this approach, the surrounding ligands are treated as point charges, and the splitting of the rare-earth ion energy levels is obtained from the CEF Hamiltonian.

$$H_{CEF} = \sum B_q^k \, O_q^k,$$

where $B_q^k$ representing the crystal-field parameters and $O_q^k$ are the Stevens operator equivalents. The crystal electric field (CEF) effects in the $R_2BaNiO_5$ series are analysed within the point-charge approximation, considering the orthorhombic crystal structure (space group Immm) with $C_{2v}$ site symmetry at the rare-earth position (0, 0, 0).

### *4.1.* $Dy_2BaNiO_5$

The $Dy^{3+}$ ions in $Dy_2BaNiO_5$ (S = 5/2, L = 5, J = 15/2) are expected to split into 2J+1=16 states, forming 8 Kramers doublets, which is described within the point-charge approximation. The obtained Stevens parameter $B_q^k$ is given in Table 1. The calculated ground state wavefunctions are: 0.885|-15/2> -0.420|-11/2> +0.192|-7/2> -0.063|-3/2> +0.008|1/2> +0.006|5/2>

+0.000|7/2> -0.006|9/2> -0.000|11/2> +0.004|13/2> +0.000|15/2>, and -0.000|-15/2> +0.004|-13/2> +0.000|-11/2> -0.006|-9/2> -0.000|-7/2> +0.006|-5/2> +0.008|-1/2> -0.063|3/2> +0.192|7/2> -0.420|11/2> +0.885|15/2>. The calculated 8 energy eigenvalues are 0, 8, 14, 21, 34, 53, 77, 102 meV, and all have a doublet eigenstate. Optical spectroscopy studies [44] reported CEF excitations at 100, 150, and 250 cm-1 (~ 12, 18, and 31 meV), which closely

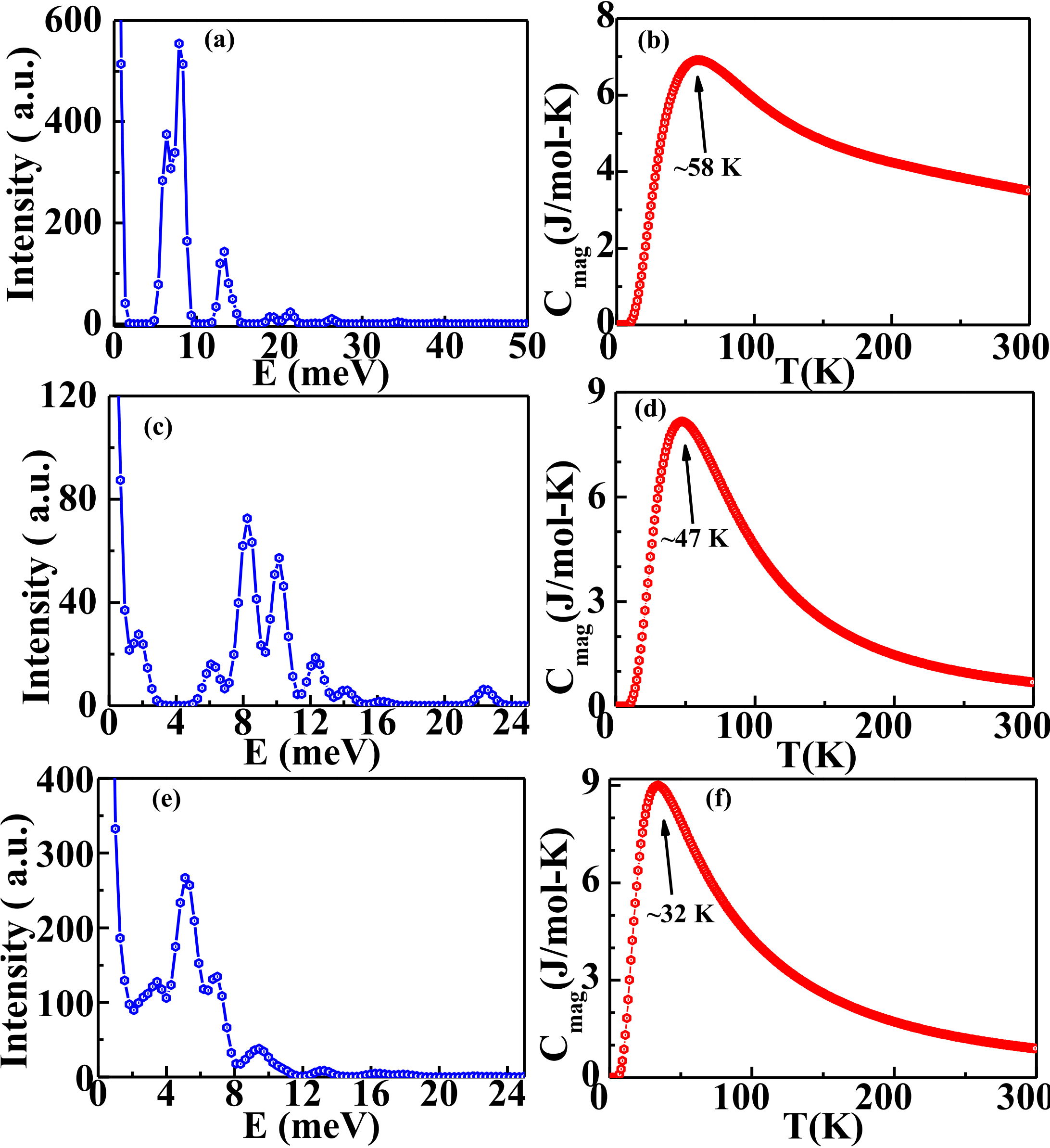

**Figure 6: Crystal electric field (CEF) excitation spectra for (a) $Dy_2BaNiO_5$, (c) $Nd_2BaNiO_5$, and (e) $Er_2BaNiO_5$, along with the corresponding magnetic heat capacity $C_{mag}$ shown in (b), (d), and (f), respectively. The $C_{mag}$ data exhibit broad Schottky type anomalies centered at approximately 58 K, 47 K, and 32 K for $Dy_2BaNiO_5$, $Nd_2BaNiO_5$, and $Er_2BaNiO_5$, respectively.**

match our point-charge calculation values of 14, 21, and 34 meV. (shown in Fig. 6a). Moreover, from this model, the magnetic heat capacity $C_{mag}$ (shown in Fig. 6b) from single ion contribution (i.e., Dy) shows a Schottky-type broad peak around 56 K, which closely matches the ordering temperature. All possible CEF transition based on the calculated CEF excitation is illustrated in Fig.7a.

**Table 1 Calculated Stevens crystal-field parameters ($B_q^k$) for $R_2BaNiO_5$ (R = $Dy^{3+}$, $Nd^{3+}$, $Er^{3+}$).**

| Parameter | $Dy^{3+}$ ($Dy_2BaNiO_5$) | $Nd^{3+}$ ($Nd_2BaNiO_5$) | $Er^{3+}$ ($Er_2BaNiO_5$) |
|---|---|---|---|
| $B_{20}$ | -0.27959 | -0.190376 | 0.119778 |
| $B_{22}$ | 0.888084 | 0.00158547 | -0.120869 |
| $B_{40}$ | -0.000285364, | -0.000244984 | 0.000163071 |
| $B_{42}$ | 0.00154542 | 0.0116454 | -0.0018402 |
| $B_{44}$ | 0.00305718 | 0.0168026 | -0.00178409 |
| $B_{60}$ | 4.38207e-07 | -3.74671e-06 | 6.54165e-07 |
| $B_{62}$ | -4.3166e-06 | 0.000389639 | -6.803e-06 |
| $B_{64}$ | 3.03414e-06 | 0.000436733 | 7.62524e-06 |

### *4.2. $Nd_2BaNiO_5$*

The $Nd^{3+}$(S=3/2, L=6, J=9/2) in $Nd_2BaNiO_5$ are expected to split into 2J+1=10 states, forming 5 Kramer's doublets. The obtained Stevens parameter $B_q^k$ is given in Table 1. The ground-state wave function consists of a mixture of |-7/2>, |-3/2>, |1/2>, |5/2>, |9/2> states, forming a Kramer's doublet with the following components, 0.001|-7/2> +0.176|-3/2> -0.004|1/2> -0.462|5/2> +0.869|9/2>, and -0.001|-7/2> -0.235|-3/2> +0.010|1/2> +0.530|5/2> -0.814|9/2>. The calculated magnetic heat capacity $C_{mag}$ from single ion contribution $Nd^{3+}$ (see Fig. 6d) shows a Schottky-type broad peak around 49K, which closely matches the ordering temperature. The calculated 5 eigenvalues are around 0, 9.3, 11.6, 18.6, and 24.5 meV (see Fig.6c). The probable CEF transitions based on the calculated CEF excitation are illustrated in Fig.7b. These results are consistent with earlier theoretical studies [45], which report anisotropic behavior in $Nd_2BaNiO_5$ arising from CEF effects. In particular, the rare-earth and Ni sublattices exhibit different preferred spin orientations, leading to competing anisotropies within the system. Such a scenario provides a natural basis for field-induced modifications of the magnetic state. Overall, the present CEF analysis highlights the importance of low-energy level splitting and anisotropy in governing the magnetic and magnetocaloric properties of $Nd_2BaNiO_5$, although the full behavior likely involves additional contributions from exchange interactions between the Nd and Ni sublattices.

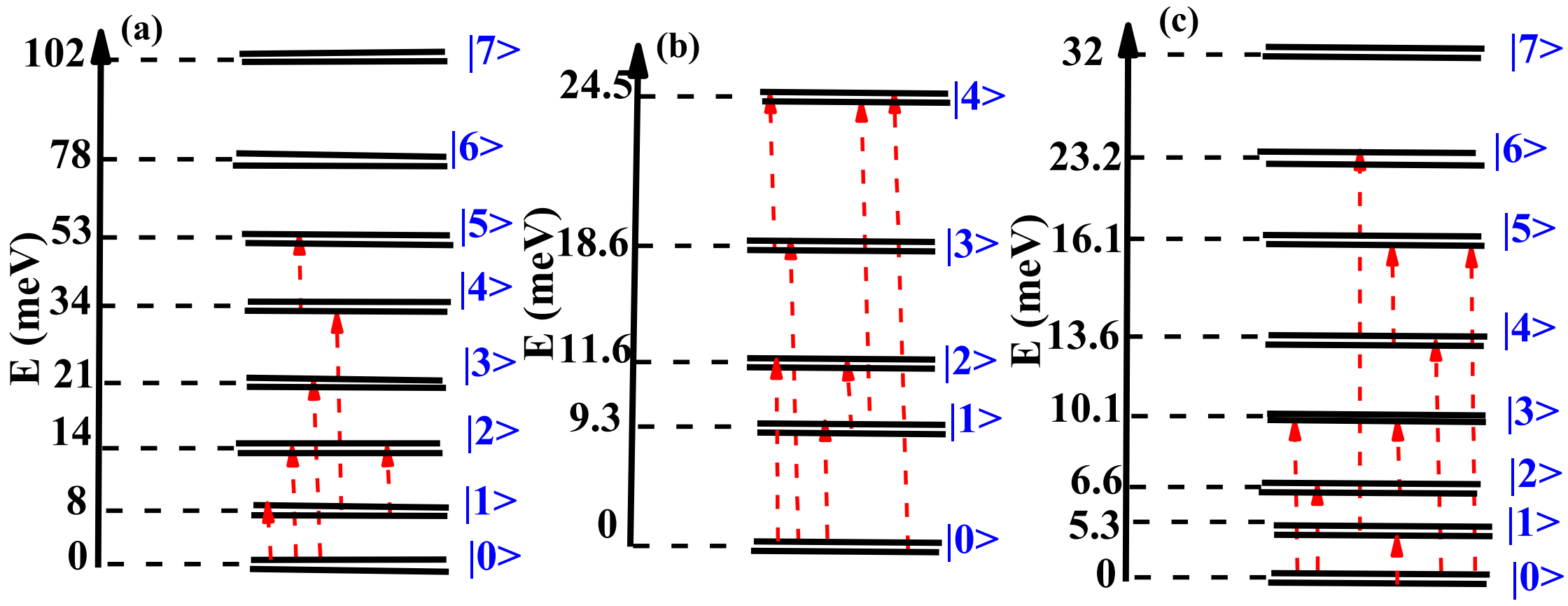

**Figure 7: Probable CEF excitation due to (a) $Dy^{3+}$ in $Dy_2NiBaO_5$ (b) $Nd^{3+}$ in $Nd_2BaNiO_5$, and (c) $Er^{3+}$ in $Er_2NiBaO_5$.**

### *4.3. $Gd_2BaNiO_5$*

For $Gd^{3+}$ ($4f^7$, L=0, S=7/2), in $Gd_2BaNiO_5$, the orbital angular momentum is quenched, resulting in an extremely weak CEF interaction. In the point-charge model scenario for $Gd^{3+}$, the J=7/2 state yields all Stevens parameters $B_q^k = 0$. So, no crystal-field splitting occurs, and as a result, the eigensystem comes out purely in the $m_J$ state with no mixing, in sharp contrast to other R-members. Hence, there is no contribution in the heat-capacity from CEF in this model, which is in sharp contrast to other R-members.

### *4.4. $Er_2BaNiO_5$*

For $Er^{3+}$ (S=6, L=3/2, J=15/2) in $Er_2BaNiO_5$, in the point charge model, the calculated Stevens' parameters are listed in Table 1. For J=15/2, the degeneracy is 2J+1=16, which splits into 8 Kramer's doublets. The ground state wave function has a mixing of |15/2>, |13/2>, |11/2>, |9/2>, |7/2>,|5/2> ,|3/2>,|1/2> that are 0.043|-15/2> +0.060|-13/2> +0.192|-11/2> +0.154|-9/2> +0.349|-7/2> +0.219|-5/2> +0.418|-3/2> +0.239|-1/2> +0.431|1/2> +0.232|3/2> +0.393|5/2> +0.194|7/2> +0.278|9/2> +0.106|11/2> +0.107|13/2> +0.024|15/2> and 0.024|-15/2> -0.107|-13/2> +0.106|-11/2> -0.278|-9/2> +0.194|-7/2> -0.393|-5/2> +0.232|-3/2> -0.431|-1/2> +0.239|1/2> -0.418|3/2> +0.219|5/2> -0.349|7/2> +0.154|9/2> -0.192|11/2> +0.060|13/2> -0.043|15/2> . The calculated CEF excitations are given in Fig. 6e. The calculated magnetic heat capacity $C_{mag}$ from single ion contribution (i.e., Er) (see Fig. 6f) shows a Schottky-type broad peak around 32 K, which closely matches the ordering temperature. The possible CEF transition based on calculated CEF excitation is illustrated in Fig.7c and the predicted transitions are 2.6 (|4>-|5>), 3.4 (|2>-|3>),5.1 (|0>-|1>),6.7 (|0>-|2>),9.4 (|0>-|3>),13.1 (|0>-|4>),16.2 (|0>-|5>),17.9 meV (|1>-|6>).

## 5. Discussion

### *5.1 Spin-orbit coupling induced magneto-caloric switching*

The $Gd^{+3}$ has almost no orbital contribution to its magnetic moment due to its half-filled 4f shell **L=0**, which results in an effective quantum number J=S=7/2**.** This leads to very low magnetic anisotropy due to negligible crystal-electric-field and spin-orbit coupling in Gd-based compounds. The $2^{nd}$-order perturbation of CEF will not have a pronounced effect. Therefore, the energy required to rotate the magnetic moments is minimal. Low anisotropy results in easier and more efficient alignment of spins with an external magnetic field, which often makes Gd a more suitable candidate for large MCE. Therefore, the CEF and SOC of other rare-earth members, Dy, Er, Nd, have an important role in magnetic anisotropy, and thereby, control the MCE more intriguingly. The absence of such conventional (positive) MCE to Inverse (negative) MCE only for $Gd_2BaNiO_5$ and multiple switching of MCE for Dy, Er and Nd members is attributed to an effect of CEF and SOC. The compelling effects of CEF, magnetic anisotropy and various exchange interactions yields magnetic phase transition and produce intriguing T and H-dependent magnetic structure due to reorientation of the spins or/and change in magnetic ground state, which governs the multiple switching of MCE in different temperature and magnetic field region. The other multiferroic 3d-4f oxide systems, e.g., $RMnO_3$ and $RMn_2O_5$ exhibit conventional MCE, [ 21,22] though, such switching from conventional MCE to inverse MCE in a 3d-4f oxide system is rarely reported. A MCE switching with negligible MCE value is documented in $GdMn_{1-x}Cr_xO_3$[46]. A few intermetallic systems, for example, $RE_2Ga_2Mg$, $R_4RhAl$, and alloys shows similar MCE switching behavior [47-49]. Neutron diffraction under magnetic field is warranted to understand the detailed spin-structure in presence of magnetic field.

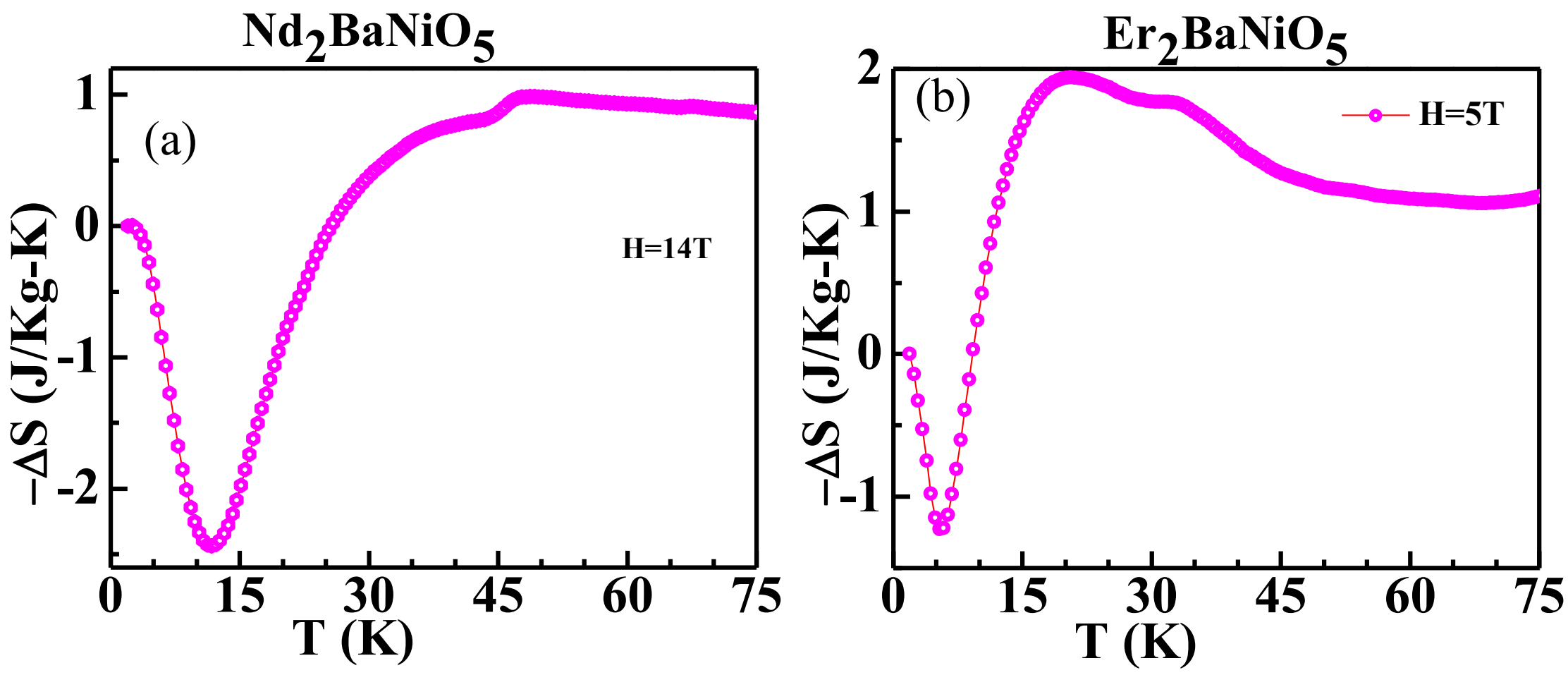

**Figure 8 Change in magnetic entropy as a function of temperature in extended region for various magnetic field.**

### *5.2 Spin-chain incipient MCE in paramagnetic region*

Here, we observe a significant MCE above the ordering temperature for all the rare-earth members in this Haldane spin-chain series – a feature that has not yet been addressed. The MCE is notable at very high temperatures for all the heavy rare-earth members, regardless of their long-range magnetic ordering, which occurs at 58 K, 32 K, and 55 K for Dy, Er, and Gd, respectively. The change in magnetic entropy for particular magnetic field is shown in Fig.8 for R= Er and Nd-members for a large temperature range upto 80 K. The presence of notable MCE at very high-temperature above $T_N$ for both heavy (say, $Er_2BaNiO_5$) and light-rare-earth member ($Nd_2BaNiO_5$) indicates the intrinsic nature of this system. The existence of short-range correlations and intriguing spin-chain effect above long-range ordering is already established in this family. [50,51]. Hence, we conclude, that the MCE above $T_N$ arises from Ni spin-chain due to short-range magnetic correlation. The low-dimensional magnetic frustration from the one-dimensional spin chain is responsible for the significant change in magnetic entropy.

**Table 2 Comparison of MCE parameters of $R_2BaNiO_5$ (R211 phase) (R = Dy, Er, Gd and Nd) and some recently reported rare-earth based cryogenic MCE materials.**

| Compounds | Corresponding temperatures (K) | $-\Delta S_M^{max}$ (field) | References |
|---|---|---|---|
| ***$Dy_2BaCuO_5$*** | 11.5 | 8.3 J/Kg-K (7T) | [24] |
| ***$Er_2BaCuO_5$*** | 7.8 | 9.6 J/Kg-K (7T) | [24] |
| ***$Ho_2BaCuO_5$*** | 12.5 | 7.03 J/Kg-K (5T) | [52] |
| ***$Er_2BaCuO_5$*** | 7.8 | 7.5 J/Kg-K (5T) | [24] |
| ***$Er_2BaZnO_5$*** | ~2.7 | 12.31 J/Kg-K (5T) | [23] |
| ***$Gd_2BaZnO_5$*** | ~2.7 | 10.28 J/Kg-K (5T) | [23] |
| ***$Dy_2BaZnO_5$*** | ~2.0 | 16.91 J/Kg-K (5T) | [23] |
| ***$Ho_2BaZnO_5$*** | ~2.0 | 15.24 J/Kg-K (5T) | [23] |
| ***$Dy_2BaNiO_5$*** | ~58.0 | 4.0 J/Kg-K (14T) | This work |
| | 18.0 | -3.1 J/Kg-K (7T) | |
| ***$Er_2BaNiO_5$*** | 22.0 | 7.8 J/Kg-K (14T) | This work |
| | 6.0 | -2.2 J/Kg-K (7T) | |
| ***$Gd_2BaNiO_5$*** | 25.0 | 6.8 J/Kg-K (14T) | This work |
| ***$Nd_2BaNiO_5$*** | 47.0 | 1.0 J/Kg-K (14T) | This work |
| | 12.0 | -2.4 J/Kg-K (14T) | |

## 6. Conclusion:

We have investigated the magnetocaloric effect and role of crystal field effect in MCE using point charge model of the Haldane-chain series for different rare-earth members with versatile properties. The heavy rare-earth member $Dy_2BaNiO_5$, exhibiting a strong spin reorientation with T and H, shows an intriguing MCE behavior with multiple switching. The other heavy-rare-earth member $Er_2BaNiO_5$ also exhibits a cross-over of magnetic entropy, attributed to the effect of a strong crystal electric field of Er-ion. The light-rare-earth members $Nd_2BaNiO_5$ show nearly similar MCE behavior to Er, related to the CEF of Nd. Interestingly, the $Gd_2BaNiO_5$, having an S-state ion (L=0), does not show any switching of MCE. A comparative summary of the magnetocaloric parameters of the present $R_2BaNiO_5$ series and recently reported rare-earth cryogenic MC materials is provided in Table.2. Our results materialize the possible control of MCE switching by tuning the CEF and SOC in the compound. We observe that magnetic entropy exists over a wide range of temperature regions for all the members in this Haldane family. Eventually, it does not vanish at high temperatures above long-range magnetic ordering. The presence of MCE in the paramagnetic regions is ascribed to spin-chain incipient short-range magnetic ordering. The study characterizes this Haldane-chain system as magnetoelectric magnetocaloric system exhibiting its potential multifunctional properties which arises from strong d-f correlations. Our interesting magnetocaloric results appeal to investigate different frustrated / spin-chain oxide systems to achieve MCE over a wide range of temperatures tuned by both Rare-earth and transition-metal ions.

**Acknowledgements:**

TB greatly acknowledges the Science and Engineering Research Board (SERB) (Project No.: SRG/2022/000044) Government of India for funding. MK acknowledges UGC for Ph.D. fellowship. Authors thanks Prof. E. V. Sampathkumaran, Tata Institute of Fundamental

Research, India, and Dr. R. Rawat, UGC-DAE Consortium for Scientific Research, Indore, India for granting the access of some experimental facilities.

**Competing interests:**

The authors declare no competing financial interests.